\documentclass[twocolumn,english,prl, showpacs,superscriptaddress]{revtex4}
\usepackage[T1]{fontenc}
\usepackage[latin9]{inputenc}
\usepackage{babel}

\usepackage{amsmath}
\usepackage{graphicx}
\usepackage{amssymb}
\usepackage{esint}
\usepackage[unicode=true,pdfusetitle,
 bookmarks=true,bookmarksnumbered=false,bookmarksopen=false,
 breaklinks=false,pdfborder={0 0 1},backref=false,colorlinks=false]
 {hyperref}

\makeatletter
\@ifundefined{textcolor}{}
{%
 \definecolor{BLACK}{gray}{0}
 \definecolor{WHITE}{gray}{1}
 \definecolor{RED}{rgb}{1,0,0}
 \definecolor{GREEN}{rgb}{0,1,0}
 \definecolor{BLUE}{rgb}{0,0,1}
 \definecolor{CYAN}{cmyk}{1,0,0,0}
 \definecolor{MAGENTA}{cmyk}{0,1,0,0}
 \definecolor{YELLOW}{cmyk}{0,0,1,0}
 }

\makeatother

\begin{document}

\title{Quantum refrigerator driven by current noise}

\author{Yi-Xin Chen}

\email{yxchen@zimp.zju.edu.cn}

\affiliation{Zhejiang Institute of Modern Physics, Zhejiang University, Hangzhou
310027, China }

\author{Sheng-Wen Li}

\affiliation{Zhejiang Institute of Modern Physics, Zhejiang University, Hangzhou
310027, China }

\affiliation{Kavli Institute for Theoretical Physics China, CAS, Beijing 100190,
China}

\begin{abstract}
We proposed a scheme to implement a self-contained quantum refrigerator
system composed of three rf-SQUID qubits, or rather, flux-biased phase
qubits. The three qubits play the roles of the target, the refrigerator
and the heat engine respectively. We provide different effective temperatures
for the three qubits, by imposing external current noises of different
strengths. The differences of effective temperatures give rise to
the flow of free energy and that drives the refrigerator system to
cool down the target. We also show that the efficiency of the system
approaches the Carnot efficiency.
\end{abstract}
\maketitle
\emph{Introduction.}--- It is an interesting problem to discuss how
small we can create a cooling machine and what would happen when quantum
effects are taken into consideration, e.g., whether a quantum refrigerator
could exceed the classical Carnot efficiency. In practice, it is also
a great challenge how to obtain lower temperature to implement quantum
tasks.

A lot of work has been done, both theoretically and experimentally
\cite{zhang_cooling_2005,Valenzuela08122006,grajcar_sisyphus_2008,wang_cooling_2009,macovei_cooling_2010,steeneken_piezoresistive_2011},
to cool a quantum system. Most of the proposals require external control
of microwave field for excitation or periodic control, and the cooling
efficiency in experiments is usually quite low. 

However, recently, Linden \emph{et al.} proposed a self-contained
refrigerator system, which has a heat engine inside to drive the whole
system \cite{linden_how_2010-1}. The system contains three qubits,
which play the roles of the target to be cooled, the refrigerator
and the heat engine, denoted by 1, 2 and 3 respectively. The three
qubits work under different temperature conditions, and they interact
through \[
\hat{V}=V_{g}(|010\rangle\langle101|+|101\rangle\langle010|).\]
 $E_{1}+E_{3}=E_{2}$ is also required to guarantee the energy conservation,
where $E_{\alpha}$ is the energy of each two-level system, $H_{\alpha}=E_{\alpha}|1\rangle_{\alpha}\langle1|$.

The first term in $\hat{V}$ describes the main cooling process of
qubit 1 (the target) by qubit 2 (the refrigerator), while the second
term describe the reversal effect. In order to make sure the cooling
process dominates, we must guarantee that the third qubit (the heat
engine) is hot enough. That is to say, the population of $|1\rangle$
of qubit 3 must be large enough, so as to enhance the cooling process.
They analyzed the problem in a phenomenological dissipation model
and demonstrated the cooling effect. Later works show that the efficiency
approaches the Carnot engine \cite{skrzypczyk_efficiency_2010,popescu_maximally_2010}.

In this paper, we propose a realizable model to implement this self-contained
quantum refrigerator by three rf-SQUID qubits, or rather, flux-biased
phase qubits. Josephson circuits techniques are relatively sophisticated
nowadays, no matter the fabrication, control or measurement, especially
when we are dealing with quantum tasks that includes not too many
qubits \cite{neeley_generation_2010,pinto_analysis_2010,bialczak_fast_2011}.
Moreover, Josephson circuits have some unique merits to build this
self-contained refrigerator system.

One of the obstacles to build this refrigerator is that, in experiments,
it is hard to build systems with 3-body interaction directly. However,
indirect 3-body interaction may arise from the transmission of basic
2-body interactions, if we impose proper detuning condition. This
is a method frequently used in optical lattice systems \cite{pachos_effective_2004}.
Fortunately, we can achieve such indirect 3-body interaction in our
system proposed in this paper.

Besides, another problem that seems more difficult is how to maintain
the three microscopic qubits in \emph{different} temperatures, which
separate from each other usually at a distance of only several microns.
We would solve it by utilizing effective temperatures of the Nyquist
noises brought in from external circuits. 

In Josephson circuits, one of the important sources of noise comes
from external currents. The current noises are unavoidably brought
in from external circuits that perform controlling or measuring, and
transfer along the circuit wires. Usually, we suppress current noise
by filters or delicate design to enhance the quantum coherence of
Josephson qubits.

In our proposal here, we pour into the three qubits with noises of
different strengths. Current noises run along wires from outside,
exactly as hot water runs along pipes from a heat source. Therefore,
equivalently, we offer the three qubits with independent different
thermal reservoirs. Moreover, the equivalent thermal reservoirs that
we provide here are stable, whose temperatures cannot be affected
by the three qubits. In experiments, Josephson circuits are usually
thermally anchored to the mixing chamber in a dilution refrigerator
at a temperature $T_{\mathrm{mix}}\approx10\,\mathrm{mK}$, while
the effective temperature caused by the current noise may be even
as high as $\sim300\,\mathrm{mK}$ \cite{grajcar_four-qubit_2006}.
We believe that we can make use of such effective temperature of noises
in our self-contained refrigerator system.

\emph{Proposal.}--- We show our circuit design in Fig.\,1. Flux bias
for each qubit is provided by external noisy current, and we assume
this is the main contribution to dissipation in our system. The three
qubits interact through mutual inductive coils overlayed together
\cite{pinto_analysis_2010,bialczak_fast_2011}, as represented in
the dashed line squre part in Fig.\,1. For simplicity, we assume
the three mutual inductive coils are identical. 

\begin{figure}
\includegraphics[width=5cm]{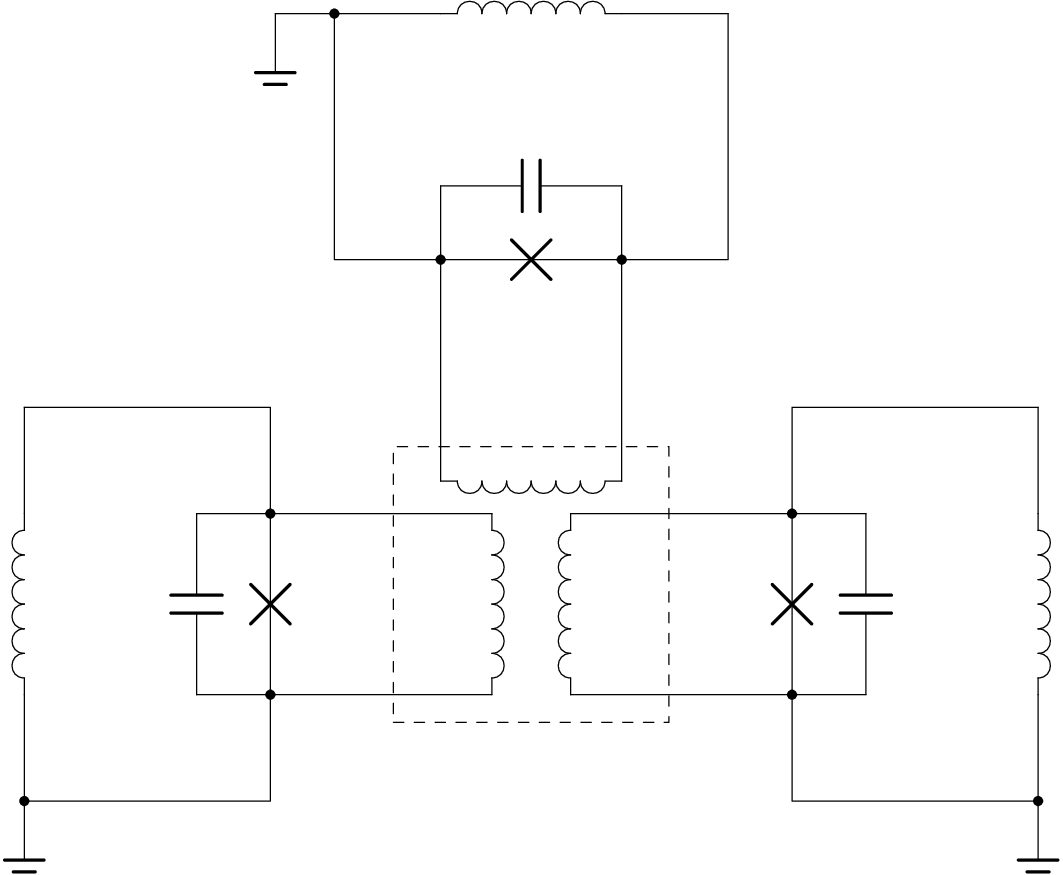}

\caption{The design of circuit of our flux-biased phase qubits system. The
part inside the dashed line square represent three identical mutual
inductance coils interacting with each other symmetrically. External
inductances provide flux biases for each qubit.}
\end{figure}

To get the quantum description of the system, we can write down the
classical equations of motion of the circuit, then get the Lagrangian
of the system, and finally, we can obtain the conjugate canonical
momentum and the Hamiltonian,

\begin{eqnarray}
\hat{H} & = & \sum_{\alpha=1}^{3}\left[\frac{\hat{Q}_{\alpha}^{2}}{2C_{\alpha}}-E_{J}^{\alpha}\cos\hat{\varphi}_{\alpha}+\frac{\tilde{\Phi}_{0}^{2}}{2L_{\alpha}}(\hat{\varphi}_{\alpha}-\Phi_{\alpha}^{\mathrm{ext}}/\tilde{\Phi}_{0})^{2}\right]\nonumber \\
 & + & \frac{\tilde{\Phi}_{0}^{2}(L_{M}+M)}{(L_{M}+2M)(L_{M}-M)}\left[\sum_{\alpha=1}^{3}\frac{1}{2}\hat{\varphi}_{\alpha}^{2}-\sum_{\alpha<\beta}\hat{\varphi}_{\alpha}\hat{\varphi}_{\beta}\right],\label{eq:H_jj}\end{eqnarray}
 where $\tilde{\Phi}_{0}=\Phi_{0}/2\pi$ and $\Phi_{0}$ is the flux
quantum. $\Phi_{\alpha}^{\mathrm{ext}}$ is the external flux imposed
to the rf-SQUID loop. $Q_{\alpha}$ is the charge carried by the capacitance
of the Josephson junction, and $\varphi_{\alpha}$ is the superconducting
phase difference across each junction. $L_{\alpha}$ is the self-inductance
of each rf-SQUID loop. $L_{M}$ and $M$ are the self and mutual inductances
of the mutual inductive coils overlayed together, as denoted in the
dashed lined square in Fig.\,1.

We set the parameters in such a way that each qubit works in a meta-stable
cubic well approximately, and choose the lowest two bound state as
a two-level system. Move the origin of $\varphi_{\alpha}$ to the
stable point of potential $U_{\alpha}(\varphi_{\alpha})$, we can
rewrite the Hamiltonian as

\begin{eqnarray}
\hat{H} & = & \sum_{\alpha=1}^{3}\hat{H}_{\alpha}+V_{\mathrm{int}}\nonumber \\
 & = & \sum_{\alpha=1}^{3}\left(\frac{p_{\alpha}^{2}}{2m_{\alpha}}+\frac{m_{\alpha}\omega_{\alpha}^{2}x_{\alpha}^{2}}{2}-\lambda x_{\alpha}^{3}\right)+g\sum_{\alpha<\beta}x_{\alpha}x_{\beta}.\label{eq:H_ho}\end{eqnarray}
 $m_{\alpha}$ , $\omega_{\alpha}$ , $\lambda$ and the coupling
constant $g$ are determined from the parameters in Eq.\,(\ref{eq:H_jj}).
$x_{\alpha}=\varphi_{\alpha}-\varphi_{\alpha}^{\mathrm{sta.}}$ is
the translated coordinate whose stable point is settled at the origin
point.

We can treat $\hat{H}_{\alpha}$ as a harmonic oscillator plus a cubic
term as the perturbation. So we have $x_{\alpha}=[\hbar/2m_{\alpha}\omega_{\alpha}]^{\frac{1}{2}}(a_{\alpha}+a_{\alpha}^{\dagger})$,
and $a_{\alpha}$ is the annihilation operator of the harmonic oscillator.
Modifications should be made to the energy and eigenstate of the corresponding
harmonic oscillator.

If we focus on the dynamics of the lowest two levels, we can write
down the low-energy effective Hamiltonian of each single qubit as
$\hat{H}_{\alpha}=E_{\alpha}|1\rangle_{\alpha}\langle1|$, where $E_{\alpha}\simeq0.95\hbar\omega_{\alpha}$
is the anharmonically modified energy of the qubit (More details of
the computation can be found in the appendix in Ref.\,\cite{pinto_analysis_2010})
.

\emph{Indirect 3-body interaction.--- }Although $V_{\mathrm{int}}$
in the Hamiltonian Eq.\,(\ref{eq:H_ho}) only involves two-body interactions,
we can still consider indirect interactions of higher order in interaction
picture, and obtain the effective Hamiltonian through \begin{eqnarray*}
\mathcal{U}_{\mathrm{int}}(t) & = & \mathbf{T}\exp\left[-\frac{i}{\hbar}\int_{0}^{t}d\tau H_{I}(\tau)\right]\\
 & = & \mathbf{1}+\frac{i}{\hbar}H_{\mathrm{eff}}t+o(t^{2}).\end{eqnarray*}
The elements of the effective Hamiltonian up to the second order is

\begin{eqnarray*}
\left\langle \vec{n}\left|H_{\mathrm{eff}}^{(1)}\right|\vec{m}\right\rangle  & = & \left\langle \vec{n}\left|V_{\mathrm{int}}\right|\vec{m}\right\rangle \equiv V_{\mathrm{int}}^{nm},\\
\left\langle \vec{n}\left|H_{\mathrm{eff}}^{(2)}\right|\vec{m}\right\rangle  & = & \sum_{k}^{E_{k}\neq E}\frac{V_{\mathrm{int}}^{nk}V_{\mathrm{int}}^{km}}{E_{k}-E},\end{eqnarray*}
 where $E_{n}=E_{m}=E$ and $|\vec{n}\rangle=|n_{1}n_{2}n_{3}\rangle$
is the state of the three qubits.

We require that $E_{1}+E_{3}=E_{2}$ and the three energy are incommensurable.
That implies that, besides the diagonal terms, the only transition
process that remains is $|010\rangle\langle101|+\mathrm{h.c.}$, when
we apply rotating-wave-approximation (RWA), i.e., when we focus on
the long-time effects. Since $V_{\mathrm{int}}$ involves only two-body
interactions, we have $\left\langle 101\left|V_{\mathrm{int}}\right|010\right\rangle =0$.
When dealing with the second order, we would come across terms like
$\left\langle n_{\alpha}\left|x_{\alpha}\right|n_{\alpha}\right\rangle $,
which would not vanish when we add anharmonic modifications to the
states. Finally, we can get an effective Hamiltonian, \begin{eqnarray}
H_{\mathrm{eff}} & = & \sum_{\alpha}D^{(\alpha)}\hat{n}_{\alpha}+\sum_{\alpha<\beta}D^{(\alpha\beta)}\hat{n}_{\alpha}\hat{n}_{\beta}+D^{(123)}\hat{n}_{1}\hat{n}_{2}\hat{n}_{3}\nonumber \\
 &  & +\tilde{g}(|010\rangle\langle101|+|101\rangle\langle010|).\end{eqnarray}
 Here, we denote $\hat{n}_{\alpha}=\left|1\rangle_{\alpha}\langle1\right|$.
The coefficients of diagonal terms $D^{(\cdots)}$ is of order $g$,
and the transition magnitude $\tilde{g}\sim g^{2}$. We can see that
the interaction strength is quite small.

\emph{Dissipation.---} Dissipation due to external current noise has
been well discussed in literature \cite{martinis_decoherence_2003,PhysRevB.71.064512,Zhou,poudel_effect_2010}.
In our system, the flux bias is provided through $\Phi_{\alpha}^{\mathrm{ext}}=L_{\alpha}^{\mathrm{ext}}\tilde{I}_{\alpha}=L_{\alpha}^{\mathrm{ext}}[I_{\alpha}+i_{\alpha}(t)]$.
$L_{\alpha}^{\mathrm{ext}}$ comes from an external coil that provides
flux bias for a single qubit, and  $i_{\alpha}(t)$ is the noise that
satisfy $\left\langle i_{\alpha}(t)\right\rangle =0$. Tracing back
to the single qubit term in the Hamiltonian of the Josephson circuit
Eq.\,(\ref{eq:H_jj}), we can figure out that the qubits interact
with the current noise through an interaction term $\gamma_{\alpha}\hat{\varphi}_{\alpha}(t)\hat{i}_{\alpha}(t)$,
where $\gamma_{\alpha}$ collects the parameters. 

The total Hamiltonian of the three qubits, each reservoir and their
interactions is

\[
\mathcal{H}_{\mathrm{SB}}=\mathcal{H}_{\mathrm{S}}+\mathcal{H}_{\mathrm{B}}+\mathcal{H}_{\mathrm{int}},\]
 where $\mathcal{H}_{\mathrm{S}}=H_{\mathrm{eff}}$, and $\mathcal{H}_{\mathrm{int}}=\sum_{\alpha}\gamma_{\alpha}\hat{x}_{\alpha}(t)\cdot\hat{i}_{\alpha}(t)$.
$\mathcal{H}_{\mathrm{B}}$ is usually treated as a collection of
harmonic oscillators phenomenologically.

We can write down the master equation of the system, after a complicated
procedure of approximations that mainly include Born-Markovian approximation
and RWA \cite{PhysRevB.71.064512,poudel_effect_2010}. The master
equation of the three-qubit system is,

\begin{eqnarray}
\partial_{t}\rho & = & -i[H_{\mathrm{eff}},\rho]+\sum_{\alpha}\mathcal{D}_{\alpha}\rho\nonumber \\
 & = & -i[H_{\mathrm{eff}},\rho]\nonumber \\
 &  & +\sum_{\alpha}\Gamma_{\alpha}\left(N_{\alpha}(E_{\alpha})+1\right)\left[2a_{\alpha}\rho a_{\alpha}^{\dagger}-\{a_{\alpha}^{\dagger}a_{\alpha},\rho\}_{+}\right]\nonumber \\
 &  & +\sum_{\alpha}\Gamma_{\alpha}N_{\alpha}(E_{\alpha})\left[2a_{\alpha}^{\dagger}\rho a_{\alpha}-\{a_{\alpha}a_{\alpha}^{\dagger},\rho\}_{+}\right].\label{eq:ME}\end{eqnarray}
 Here, $N_{\alpha}(E_{\alpha})=[\exp(\beta_{\alpha}E_{\alpha})-1]^{-1}$.
$a_{\alpha}^{\dagger}$ and $a_{\alpha}$ are the correponding harmonic
operators of each qubit. $\Gamma_{\alpha}N_{\alpha}(E_{\alpha})\propto S_{I}^{\alpha}(E_{\alpha}/\hbar)$,
where $S_{I}^{\alpha}(\omega)$ is the power spectral of current noise
reflecting the dissipative impedance of the circuit, \[
S_{I}^{\alpha}(\omega)=\int_{0}^{\infty}e^{-i\omega t}\left\langle i_{\alpha}(t)i_{\alpha}(0)\right\rangle dt.\]

\emph{Steady solution.}--- We concentrate on the equilibrium behaviour
of the system here, and especially, we want to obtain the final steady
distribution $\left\langle \hat{n}_{\alpha}\right\rangle $ of each
single qubit. Then we can get the effective temperatures. 

We can get eight independent linear equations about $\left\langle \hat{n}_{\alpha}\right\rangle $,
$\left\langle \hat{n}_{\alpha}\hat{n}_{\beta}\right\rangle $, $\left\langle \hat{n}_{1}\hat{n}_{2}\hat{n}_{3}\right\rangle $
and $\left\langle \Delta v\right\rangle \equiv\tilde{g}\left\langle a_{1}a_{2}^{\dagger}a_{3}\right\rangle -\mathrm{h.c.}$,
by multiplying respective observables to the master equation and then
tracing out the average. And we also have the steady solution,

\begin{equation}
\left\langle \hat{n}_{\alpha}\right\rangle =\frac{N_{\alpha}}{2N_{\alpha}+1}+\frac{(-1)^{\alpha+1}i\left\langle \Delta v\right\rangle }{2\Gamma_{\alpha}(2N_{\alpha}+1)},\label{eq:n}\end{equation}
 where $N_{\alpha}=N_{\alpha}(E_{\alpha})$. Let $\mathcal{M}_{\alpha}=\Gamma_{\alpha}(2N_{\alpha}+1)$,
we have \begin{eqnarray*}
\left\langle \Delta v\right\rangle  & = & \frac{i\tilde{g}^{2}G}{X_{1}+\tilde{g}^{2}(X_{2}+X_{3})}(N_{1}N_{3}-N_{2}-N_{1}N_{2}-N_{2}N_{3})\\
 & \equiv & i\xi(N_{1}N_{3}-N_{2}-N_{1}N_{2}-N_{2}N_{3}),\end{eqnarray*}
 and \begin{eqnarray*}
G & = & 4\Gamma_{1}\Gamma_{2}\Gamma_{3}(\mathcal{M}_{1}+\mathcal{M}_{2}+\mathcal{M}_{3})\prod_{\alpha<\beta}(\mathcal{M}_{\alpha}+\mathcal{M}_{\beta}),\\
X_{1} & = & 2(1+\frac{A^{2}}{B^{2}})\mathcal{M}_{1}\mathcal{M}_{2}\mathcal{M}_{3}\left(\mathcal{M}_{1}+\mathcal{M}_{2}+\mathcal{M}_{3}\right)^{2}\\
 &  & \times\prod_{\alpha<\beta}(\mathcal{M}_{\alpha}+\mathcal{M}_{\beta}),\\
X_{2} & = & \left[4\mathcal{M}_{1}\mathcal{M}_{2}\mathcal{M}_{3}+\sum_{\alpha<\beta}\mathcal{M}_{\alpha}\mathcal{M}_{\beta}(\mathcal{M}_{\alpha}+\mathcal{M}_{\beta})\right]\\
 &  & \times\prod_{\alpha<\beta}(\mathcal{M}_{\alpha}+\mathcal{M}_{\beta}),\\
X_{3} & = & -\Gamma_{1}\Gamma_{2}\mathcal{M}_{1}\mathcal{M}_{2}(\mathcal{M}_{1}+\mathcal{M}_{2})(\mathcal{M}_{1}+\mathcal{M}_{2}+2\mathcal{M}_{3})\\
 &  & +\Gamma_{1}\Gamma_{3}\mathcal{M}_{1}\mathcal{M}_{3}(\mathcal{M}_{1}+\mathcal{M}_{3})(\mathcal{M}_{1}+2\mathcal{M}_{2}+\mathcal{M}_{3})\\
 &  & -\Gamma_{2}\Gamma_{3}\mathcal{M}_{2}\mathcal{M}_{3}(\mathcal{M}_{2}+\mathcal{M}_{3})(2\mathcal{M}_{1}+\mathcal{M}_{2}+\mathcal{M}_{3}),\end{eqnarray*}
 where $B=\sum_{\alpha}\mathcal{M}_{\alpha}$, and $A=D^{(1)}-D^{(2)}+D^{(3)}+D^{(13)}$.

The steady solution of $\left\langle \hat{n}_{\alpha}\right\rangle $
contains two terms. When there is no interaction between the three
qubits, they would respectively decay into each Boltzmann distribution,
resulting from the weak coupling with their reservoirs, as represented
by the first term in Eq.\,(\ref{eq:n}). The second term is the contribution
of interactions between qubits and that alter the populations. The
effective temperatures would be changed as long as $\left\langle \Delta v\right\rangle \neq0$.

\emph{Cooling condition and efficiency.--- }When we say that we want
to make the system run as a refrigerator in order to cool down qubit
$1$, we have actually implied that the initial temperature of qubit
1 is the lowest, otherwise, it could be cooled through heat transport.
We can check that each term in $\xi$ is positive. Therefore, to lower
the population number $\left\langle \hat{n}_{1}\right\rangle $, we
must make sure that $i\left\langle \Delta v\right\rangle <0$, and
that gives the cooling condition,

\begin{align*}
(\beta_{1}-\beta_{2})E_{1} & <(\beta_{2}-\beta_{3})E_{3},\\
\mathrm{or,\qquad}\frac{E_{1}}{E_{3}} & <\frac{\beta_{2}-\beta_{3}}{\beta_{1}-\beta_{2}}.\end{align*}
Remember that $E_{1}+E_{3}=E_{2}$. As we know that $T_{1}$ should
be the lowest, so we must have $T_{1}<T_{2}<T_{3}$.

In order to compute the cooling efficiency of the refrigerator, we
have to compare the amounts of heat exchange of the target (qubit
$1$) and the heat engine (qubit $3$) with their environments. We
should go back to the master equation Eq.\,(\ref{eq:ME}). The unitary
term represents the contribution of doing works between the three
qubits, and the second dissipative one represents the heat exchange
with the environment. Therefore, we can get the heat exchange of each
qubit with their environments per unit time by \[
Q_{\alpha}=\mathrm{Tr}\left[E_{\alpha}\hat{n}_{\alpha}\cdot\mathcal{D}_{\alpha}\rho\right]=(-1)^{\alpha}i\left\langle \Delta v\right\rangle E_{\alpha}.\]

Now we arrive at an interesting result that the efficiency of the
system is \[
\eta^{Q}=\frac{Q_{C}}{Q_{H}}=\frac{E_{1}}{E_{3}}<\frac{1-\frac{T_{R}}{T_{H}}}{\frac{T_{R}}{T_{C}}-1}\equiv\eta_{\mathrm{max}}^{Q}.\]
 This is exactly same with the results in previous work in Ref.\,\cite{skrzypczyk_efficiency_2010}.
In their work, they also figure out that this is the upper bound on
the efficiency of any such engine running between three reservoirs
which extracts heat from the bath at $T_{C}$ using a supply of heat
from the bath at $T_{H}$.

We have to emphasize that $T_{\alpha}$ here is the temperature of
the reservoir where each qubit stays, and it is completely determined
by the strength of noise from external circuit. The change of the
population of the qubits does not affect the temperature of reservoirs,
which can be only controlled through external filter circuits. The
thermal temperature of the qubits provided by the dilution refrigerator
is the same, while the strengths of random motions of the electrons
inside the conducting wires are different between the three qubits,
and that give rise to an equivalent effect of providing different
thermal conditions.

In Ref.\,\cite{linden_how_2010-1}, it was mentioned that there is
no other theoretical limitation for the refrigerator even when approaching
the absolute zero. However, in our system implemented by Josephson
circuit proposed here, we have some additional constraints to the
working conditions. First, the temperature of the heat engine $T_{H}$
must not be too high comparing with the excitation energy $E_{3}\sim10\,\mathrm{GHz}\sim1\,\mathrm{K}$,
otherwise, population of higher energy levels must be taken into account,
and we cannot treat our system simply as a two-level system. Second,
if we greatly suppress the external noise of qubit $1$ in order to
make $T_{C}$ lower ($\sim50\,\mathrm{mK}$), the low frequency $1/f$
noise would dominate in the dissipation, therefore, our analysis above
based on Markovian approximation in invalid. After all, it has to
be tested in realistic experiments how cold the refrigerator proposed
here could achieve. However, this could still provide an astonishingly
high efficiency when we compare it with macroscopic classical heat
machines. 

\emph{Summary.--- }In this paper, we proposed a Josephson circuit
system to implement a self-contained refrigerator. We demonstrated
the configuration of the circuit, analyzed the dynamics of the system
and got the final steady distribution. By controlling the different
strengths of the noises pouring into the qubits, surely the refrigerator
would cool down the target with high efficiency even approaching up
to the Carnot up-bound. We believe our proposal is realizable with
present technology.

The system contains a microscopic heat engine to maintain the running
of the whole system automatically. The power actually comes from the
external circuit noises. We do not need direct control of external
control such as periodic microwave fields. What surprised us is that
such a microscopic quantum heat machine driven by noises could run
with quite high efficiency. This gives us an inspiration that maybe
we can develop more quantum devices driven by noises. No need to say,
there is a great advantage when there is a self-contained energy provider
in microscopic quantum tasks. What's more, noises of many types may
be utilized, such as noises from external current, flux or light,
maybe even from inner fabrication defects.

When our work was almost completed, we noted that Mari and Eisert
proposed a similar mechanism of cooling by heating in an opto-mechanical
system\,\cite{mari_cooling_2011}.

\emph{Acknowledgments.--- }The work is supported in part by the NSF
of China Grant No. 10775116, No. 11075138, and 973-Program Grant No.
2005CB724508. S.-W. Li would like to thank Heng Fan and Jian Ma for
helpful suggestions.

\bibliographystyle{apsrev}

\end{document}